\documentclass[preprint]{aastex}

\usepackage{graphicx}

\shorttitle{An Enigmatic Pointlike Feature within the HD 169142 Transitional Disk}
\shortauthors{Biller et al.}
\slugcomment{Accepted to ApJ Letters}

\begin{document}


\title{An Enigmatic Pointlike Feature within the HD 169142 Transitional Disk}


\author{Beth A. Biller\altaffilmark{1, 2}, Jared Males\altaffilmark{3,*},
  Timothy Rodigas\altaffilmark{4},  Katie Morzinski\altaffilmark{3,*}, 
  Laird M. Close\altaffilmark{3}, Attila Juh\'asz\altaffilmark{5},
  Katherine B. Follette\altaffilmark{3},
Sylvestre Lacour\altaffilmark{6}, Myriam Benisty\altaffilmark{7},
Aurora Sicilia-Aguilar\altaffilmark{8,9}, Philip M.
Hinz\altaffilmark{3}, Alycia Weinberger\altaffilmark{4},  
Thomas Henning\altaffilmark{2}, J\"org-Uwe Pott\altaffilmark{2}, 
Micka\"el Bonnefoy\altaffilmark{10}, Rainer K\"ohler\altaffilmark{2}}      

\altaffiltext{1}{Institute for Astronomy, University of Edinburgh,
  Blackford Hill, Edinburgh, UK}
\altaffiltext{2}{Max-Planck-Institut f\"ur Astronomie, K\"onigstuhl 17, 69117 Heidelberg, Germany}
\altaffiltext{3}{Steward Observatory, University of Arizona, 933,
  N. Cherry Ave, Tucson, AZ, 85719, USA}
\altaffiltext{*}{NASA Sagan Fellow}
\altaffiltext{4}{Department of Terrestrial Magnetism, 
Carnegie Institution of Washington, 5241 Broad Branch Road, NW
Washington, DC 20015, USA}
\altaffiltext{5}{Leiden Observatory, Leiden University, P.O. Box 9513, 2300 RA Leiden, The Netherlands}
\altaffiltext{6}{LESIA, CNRS/UMR-8109, Observatoire de Paris, UPMC, Universit\'e Paris Diderot, 5 place Jules Janssen, 92195 Meudon, France}
\altaffiltext{7}{UJF-Grenoble 1 / CNRS-INSU, Institut de Plan\'etologie et d'Astrophysique de Grenoble (IPAG) UMR 5274,
Grenoble, F-38041, France}
\altaffiltext{8}{Departamento de F\'{\i}sica Te\'{o}rica, Facultad de Ciencias, Universidad Aut\'{o}noma de Madrid, 28049 Cantoblanco, Madrid, Spain}
\altaffiltext{9}{SUPA, School of Physics and Astronomy, University of St Andrews, North
Haugh, St Andrews KY16 9SS, UK}
\altaffiltext{10}{Universit\'{e} Grenoble Alpes, IPAG, F-38000 Grenoble, France. CNRS, IPAG, F-38000 Grenoble, France}


\email{bb@roe.ac.uk}






\begin{abstract}
We report the detection of a faint pointlike feature possibly related
to ongoing planet-formation in the disk of the 
transition disk star HD 169142.  The pointlike feature has a
$\Delta$mag(L)$\sim$6.4, at a separation of $\sim$0.11$\arcsec$ and
PA$\sim$0$^{\circ}$.  Given its lack of an H or K$_{S}$ counterpart despite its relative
brightness, this candidate cannot be explained by purely photospheric
emission and must be a disk feature heated by an as yet unknown source.
Its extremely red colors make it highly unlikely to be a background
object, but future multi-wavelength followup is necessary for
confirmation and characterization of this feature.
\footnote{Based on observations made with ESO Telescopes at the La
  Silla Paranal Observatory under programme ID 091.C-0572.}
\footnote{This letter includes data gathered with the 6.5 Magellan
  Telescopes located at Las Campanas Observatory (LCO), Chile}
\end{abstract}





\section{Introduction}
Transition disks trace a key step in the formation of planetary systems,
intermediate between gas-rich protoplanetary disks and debris disks, where
primordial gas is cleared away, leaving only remnant dust.  
These disks are observationally identified by weak mid-IR
emission (at $\sim$15 $\mu$m) relative to the Taurus median spectral
energy distribution (Najita et al. 2007, i.e. the median SED of
primordial disks in the young ($<$2 Myr) Taurus star-forming region).
While numerous physical processes may be responsible for the depletion
of gas and dust in transition disks, cleared gaps in particular may be
an indicator of a planet or brown dwarf companion in the midst of
formation.  Thus, these disk systems have been key targets for direct
imaging searches for planets.

Transition disks tend to be found in young star-forming 
regions $>$100 pc from Earth.  Only recently have imaging techniques been
available to probe the inner portions of these disks, where planets
are likely to form.  In the last few years,
there have been a number of notable discoveries of companions in
transition disks with cleared gaps.  Via the interferometric
sparse-aperture masking (henceforth SAM) technique \citep[]{Lac11b}, 
\citet[]{Kra12} found LkCa 15b, a candidate protoplanet embedded 
within the disk of a $\sim$2 Myr solar analog.  \citet[]{Bil12} 
detected a 0.1-0.4 M$_{\Sun}$ companion in the disk around the
Herbig Ae/Be star HD 142527.  This binary companion was recently confirmed 
and shown to be 
accreting via direct imaging \citep[]{Clo13}.  HD 142527 has a particularly 
wide gap in its disk ($>10$-120 AU, Fukugawa et al. 2006, Rameau et
al. 2012, Casassus et al. 2012), which may be explained by a stellar 
companion on an eccentric orbit.  Finally, \citet[]{Qua13a} reported
an as yet unconfirmed candidate protoplanet detection in the disk of the Herbig 
Ae/Be star HD 100546.

The Herbig Ae/Be star HD 169142 possesses a nearly face-on transition
disk, and has been studied in detail both spectroscopically  (Meeus et
al. 2001, van Boekel et al. 2005), and through imaging (Habart et al. 2006,
Kuhn et al. 2001, Hales et al. 2006, Grady et al. 2007, Fukugawa et
al. 2010, Honda et al. 2012, Mari{\~n}as et al. 2011).
Recently, \citet[]{Qua13b} detected a well-resolved annular gap from 40 - 70 AU via
polarized light imaging.  Thus, like LkCa 15, HD
142527, and HD 100546, it is an excellent candidate to 
possess a planetary mass or brown dwarf companion in the midst of
formation.  

\section{Stellar Parameters}

Adopted stellar parameters are presented in
Table~\ref{tab:properties}.  In general, we adopt the stellar
parameters from Table 1 of Quanz et al. 2013b.  An accurate age
estimate is particularly important for estimating candidate companion
properties; HD 169142 has been assigned a fairly wide range of ages.
Guimar{\~a}es et al. 2006 assign a comparatively young age of 1-5 Myr
based in HD 169142's HR diagram position, while Blondel $\&$ Djie 2006
claim a similar mass and age as $\beta$ Pic for HD 169142.  Grady et
al. 2007 find a comoving companion 9.3'' separated from HD 169142.
This object is a 130 mas separation weak-line T Tauri binary, thus
they assign an age of 6$^{+6}_{-3}$ Myr for the entire
system.  For our analysis here, we thus adopt an age range of 3-12
Myr.  Sylvester et al. 1996 derive a photometric distance of 145 pc to
HD 169142, based on an A5 spectral type; Blondell \& Djie 2006 update
this to 151 pc based on a spectral type of A7V.  As derived spectral
type will strongly affect the photometric distance, we adopt a
distance range of 145$\pm$10 pc here to account for any uncertainty in
spectral typing.

\section{Observations and Data Reduction}

\subsection{July 2013 VLT NACO Vortex Observations}

First epoch observations were conducted using the novel annular groove
phase mask (henceforth AGPM) vector
vortex coronagraph \citep[]{Maw13} with the NACO camera at the VLT
\citep[]{Len03, Rou03}.  The AGPM coronagraph uses an annular groove
phase mask to redirect on-axis starlight out of the
pupil \citep[]{Maw05}.  In this manner, the AGPM coronagraph enables
L' band contrasts of $\Delta$mag$>$7.5 mag at inner working angles 
down to 0.09$\arcsec$ \citep[]{Maw13}.  HD 169142 was observed 
in L' from 02:47 UT to 04:54 UT on 14 July 2013, covering 
nearly an hour both before and after transit.  The derotator was
turned off to enable azimuthal differential imaging (ADI) techniques.
Seeing varied from 1-2$\arcsec$ over the observation.  We used a base 
exposure time of 0.25 s with 120 coadds and obtained a total of 1 hour 15
minutes on-sky exposure time.  Sky frames were interspersed after every 20 frames
and 0.05 s unsaturated exposures were taken at the beginning of the sequence.
Care was taken to keep the primary star centered under the AGPM.
Observation details are presented in Table~\ref{tab:observations}.

Data were flat-fielded and bad-pixel corrected using dome flats 
and darks taken as part of ESO standard calibrations.
Numerous dust spots are apparent on the AGPM, thus, slight
misalignments between science and sky frames can produce significant 
cosmetic errors.  Thus, we carefully selected the sky frame which
minimized cosmetic errors for each science frame.  Eleven science frames
(out of 126 total science frames) 
which showed significant cosmetic errors even after sky subtraction
were discarded.  Each frame was centroided on the raw data using the 
idl routine mpfit2dpeak. 

After basic data reduction and centroiding on the center of the
coronagraphic mask using the IDL mpfit2dpeak routine, we analyze the data using 3 
independent principal component analysis (PCA) pipelines
 (following the algorithms of Soummer et al. 2012, Amara \& Quanz 2012).  
All three pipelines yielded similar results -- we report here the
results using the publically available pipeline of Dimitri Mawet 
(http://www.sc.eso.org/$\sim$dmawet/DimitriMawet/IDL\_PCA\_pipeline.html).  

\subsection{April 2014 Magellan AO Observations}


Followup observations were conducted using the 585 actuator 1000 Hz
adaptive secondary AO System (MagAO) at the 6.5m Magellan Clay
Telescope (Close et al. 2013).  With 378 corrected modes, the MagAO
system is one of the highest sampled AO systems on a large telescope
($>$5 m) and has demonstrated high spatial resolutions (20-30 mas) in the visible 
(as short as $\lambda$ =0.6 $\mu$m, Close et al. 2013). MagAO feeds 
both a visible (VisAO, 0.5-1 $\mu$m) and infrared camera (Clio2, 
1-5 $\mu$m) simultaneously. 

We obtained a total of 5 datasets for HD 169142 with MagAO from 8-15
April 2014.  Observing details are presented in
Table~\ref{tab:observations}.  Four of these datasets were obtained
with the Clio2 Narrow camera including two H band observations, one
K$_{S}$ band observation, and one 3.9 $\mu$m 5$\%$ filter observation.
We also obtained a deep z' dataset using the VisAO camera.  The
derotator was turned off to enable angular differential imaging
(ADI) techniques.  All MagAO data was reduced in the standard manner
(flat-field correction, sky-subtraction, bad-pixel correction).  The
H, K$_{S}$, and z' datasets remained unsaturated in the core, thus we align
our images using the unsaturated core.  We then process our data
using the independent PCA-based pipelines of CoIs Rodigas and Males
(see e.g. Rodigas et al. 2014 and Males et al. 2014).
Both pipelines yielded similar results.

At 3.9 $\mu$m, we acquired data for HD 169142 as well as for a
PSF star of similar magnitude.  We achieved very high Strehl ratios
for both stars in dry, photometric conditions.  Seeing remained around 
0.4-0.6$\arcsec$ for the entire night.  After image registration, we built a PSF from the
nearby star images and subtracted the PSF from each of our MagAO/Clio2
images of HD 169142, then rotated to place North up/East left.  Our best
subtraction is for a PSF scaling of 115$\%$, which removes the Airy
rings quite well.  Dark pixels in the core
are due to saturation/non-linearity.

\section{Results}

\begin{figure}
\includegraphics[width=5in]{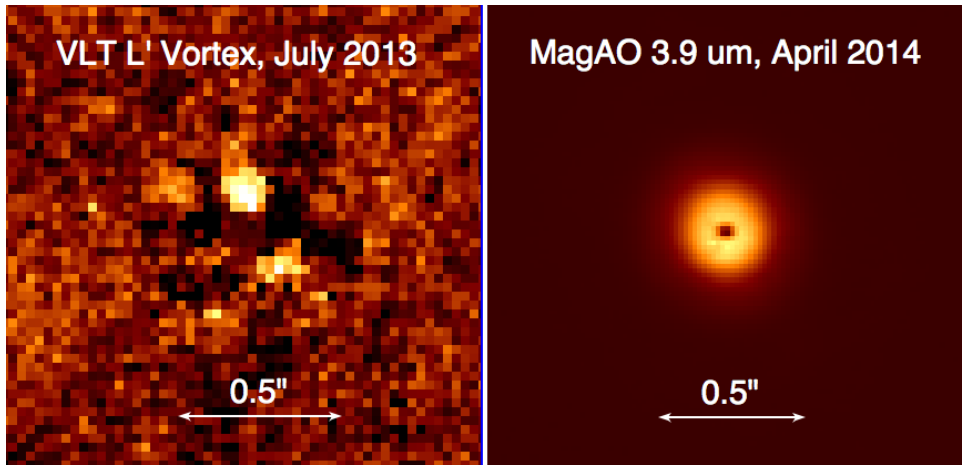}
\caption{
 Left: L' July 2013 NACO Vortex
image.  We find a pointlike feature in the disk with
$\Delta$mag=6.4$\pm$0.2, 
at a separation of 0.11$\pm$0.03$\arcsec$ and PA of
0$\pm$14$^{\circ}$.
 Right: 3.9 $\mu$m April 2014 PSF-subtracted MagAO CLIO-2 image.
Dark pixels in the core are due to saturation/non-linearity.
The April 2014 MagAO image reached shallower contrasts 
than the original vortex image, hence the faint point-like feature
is not retrieved.
Our noncoronagraphic dataset was too shallow to retrieve the candidate 
($\Delta$mag=5.6 at 0.11'').
\label{fig:image1}}
\end{figure}

\begin{figure}
\begin{tabular}{cc}
\includegraphics[width=2in, angle=90]{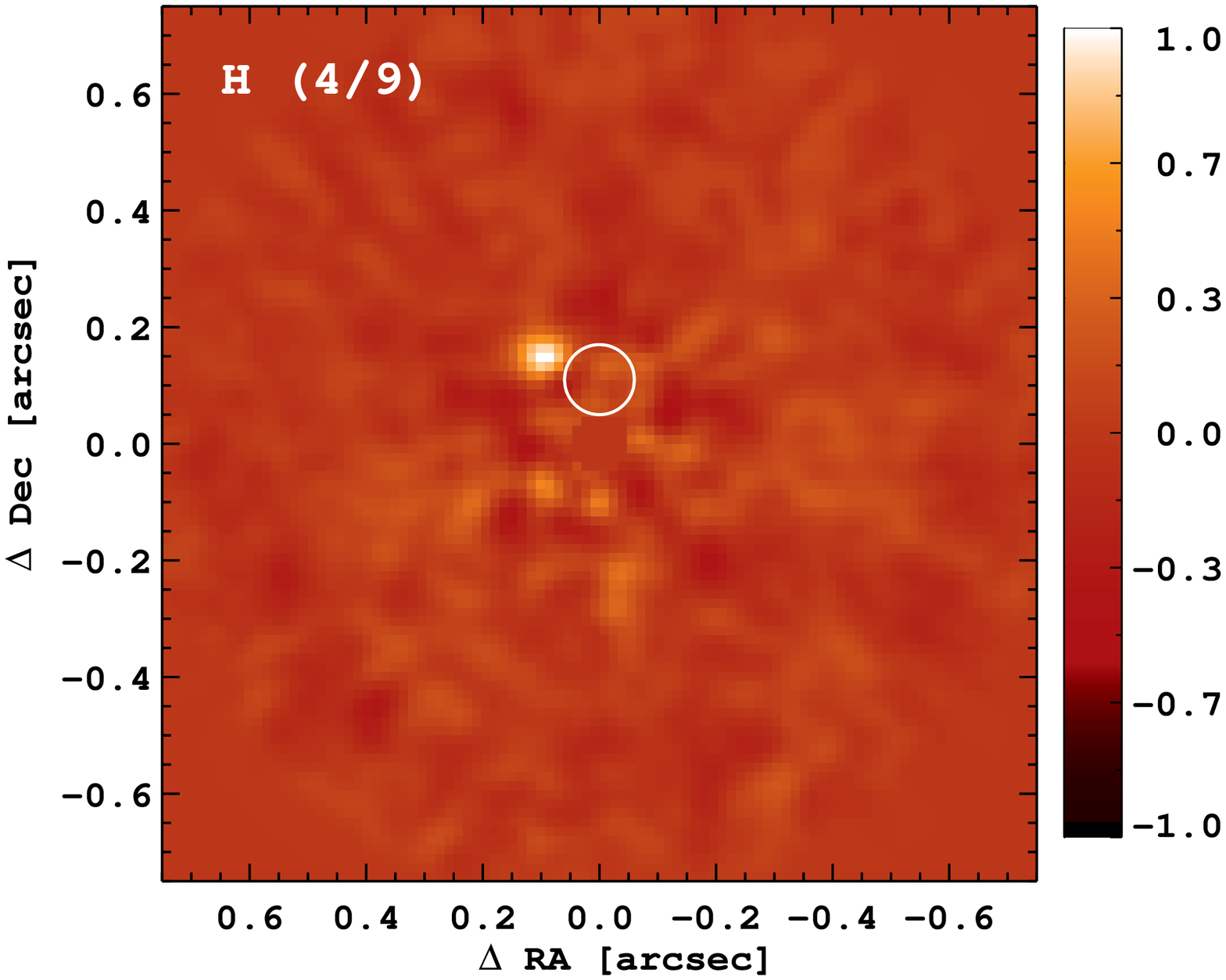} & 
\includegraphics[width=2in, angle=90]{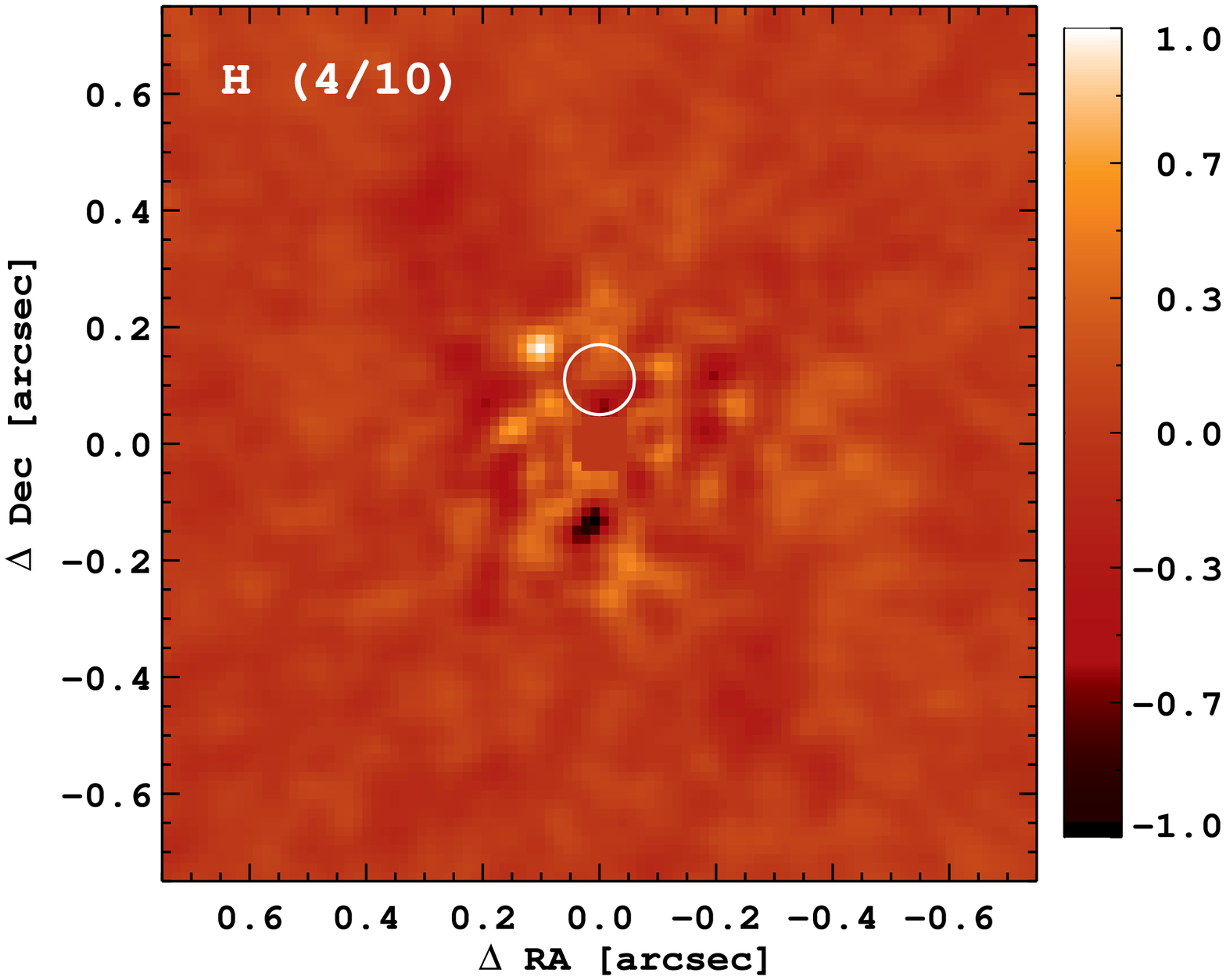} \\
\includegraphics[width=2in, angle=90]{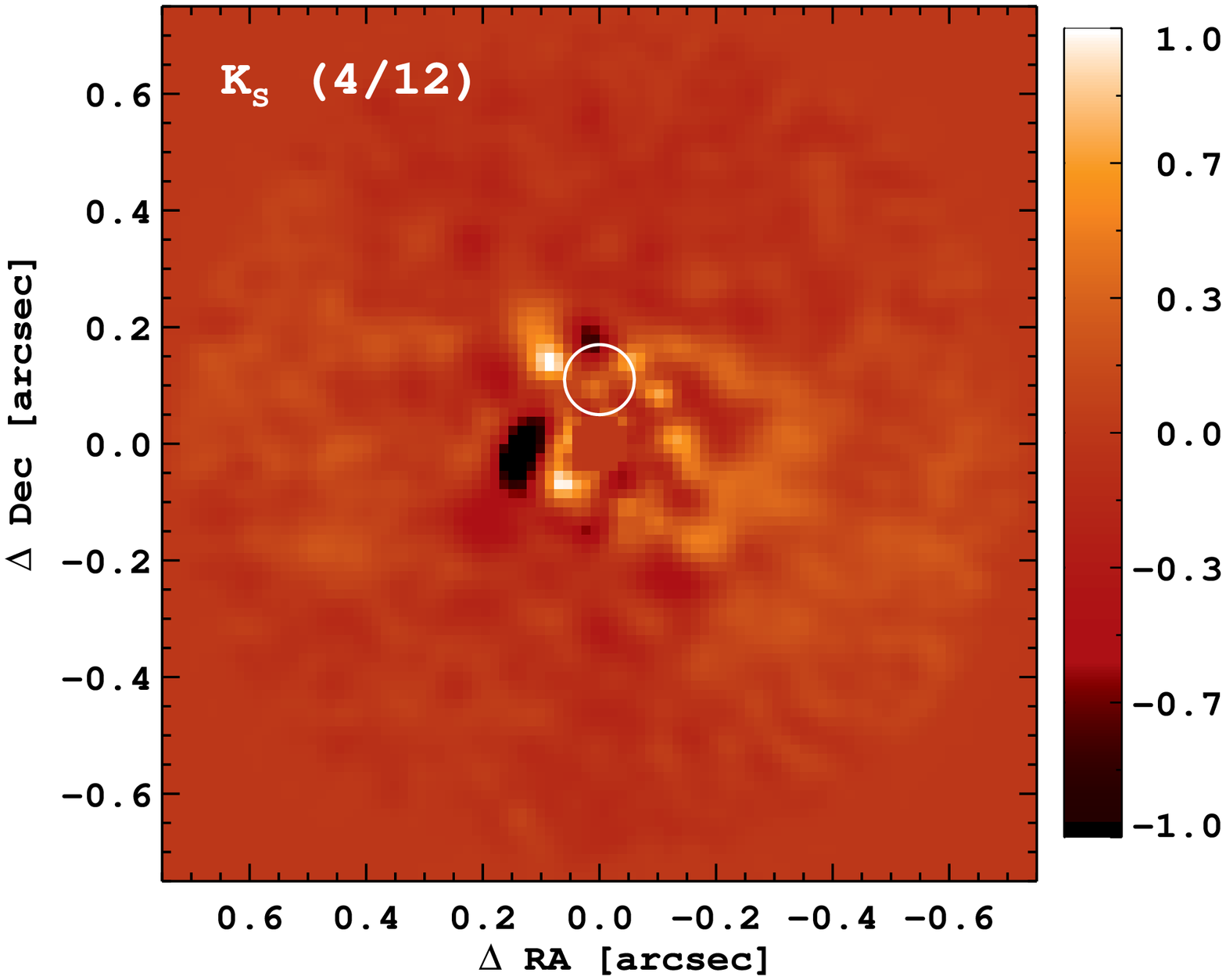} & 
\includegraphics[width=2in, angle=90]{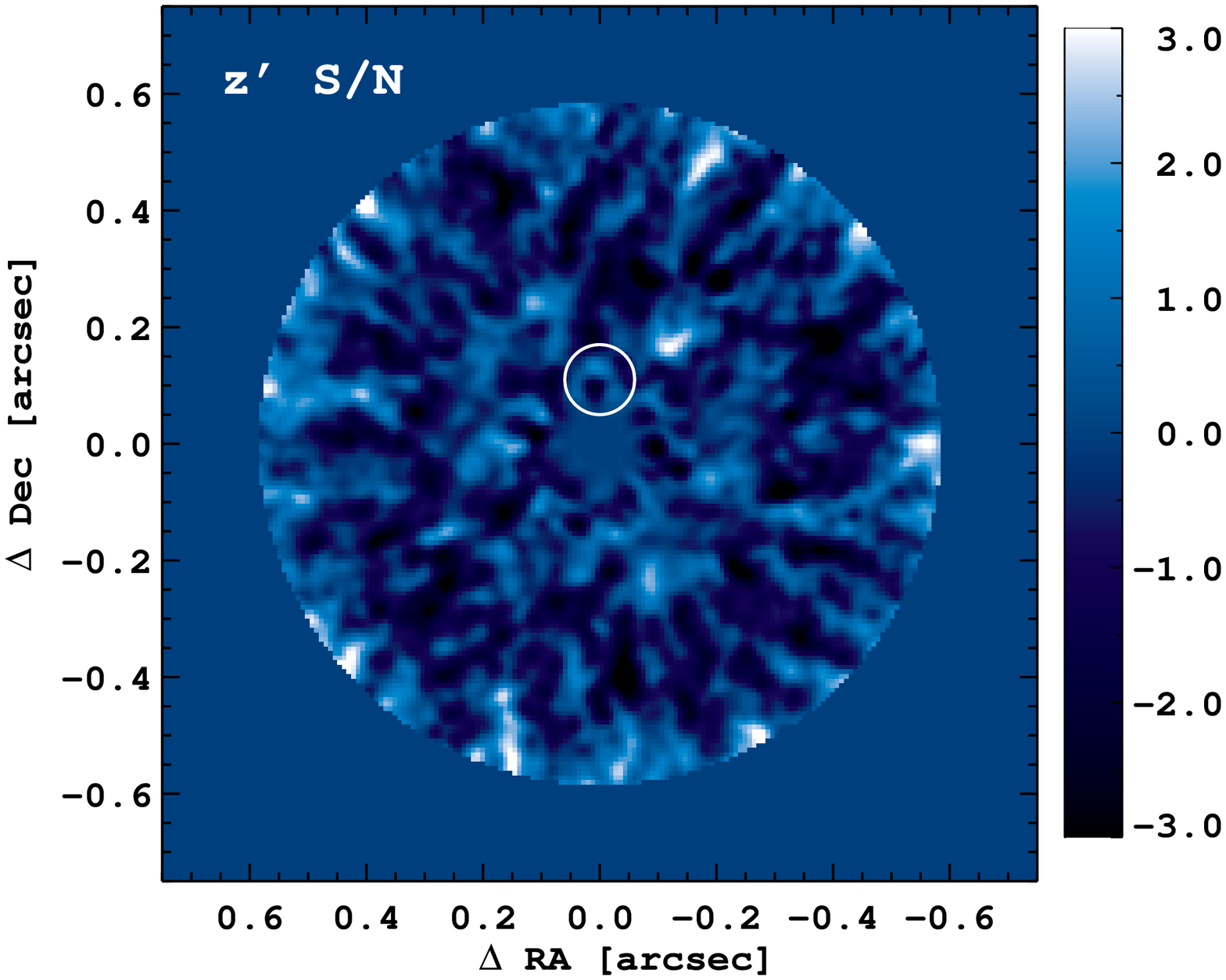} \\
\end{tabular}
\caption{
Top left and
right : H band April 2014 MagAO CLIO-2 images over two consecutive nights,
bottom left: K band April 2014 MagAO CLIO-2 image, bottom right: z' band
April 2014 MagAO VisAO image.  All images have been reduced with PCA. 
The position of the L' pointlike feature is circled.
\label{fig:image2}}
\end{figure}

In July 2013, we detected a faint, pointlike feature in our L' vortex
dataset at $\sim$0$^{\circ}$ PA, with a separation of $\sim$0.11".   
This pointlike feature was independently imaged in June 2013 by Reggiani et al. 2014.
Multiband followup observations were conducted using 
Magellan AO in April 2014.  Images from all epochs are presented in
Figs.~\ref{fig:image1} and ~\ref{fig:image2}.
Astrometry and photometry are presented in Table~\ref{tab:properties}.
The correct rotation to place north up and east left was verified at
each epoch.  For the July 2013 vortex data, a dataset on HD 142527 was
acquired on the same run.  The bright HD 142527 disk (see e.g.
Fukagawa et al. 2006, Rameau et al. 2012) is correctly derotated by
the pipeline.  For the April 2014 MagAO data, the correct derotation
has been validated using images of the Trapezium cluster.

\subsection{July 2013 VLT NACO Vortex Observations}

L' photometry and astrometry was derived for the pointlike feature by inserting 
scaled PSF images into the raw data and selecting the fluxes
and positions that best match the actual feature. 
Best astrometry and photometry is presented in Table~\ref{tab:properties}.
Assuming a point source, our best subtraction yielded $\Delta$mag=6.4$\pm$0.2
for this feature, at a separation of 0.11$\pm$0.03$\arcsec$ and PA of
0$\pm$14$^{\circ}$.  Given that the inner working angle of the vortex
coronagraph is 0.09$\arcsec$ \citep[][]{Maw13}, this pointlike feature is
marginally resolved at best.

To estimate the S/N level of our detection, 
we calculated the mean and standard deviation in a circular
aperture with diameter of $\lambda$/D centered on the pointlike feature and
in the 6 additional resolution elements available at this separation
and then calculated S/N from equation 9 of \citet[][]{Maw14}
(appropriate for speckle-dominated regions): 

\begin{equation}
S/N = \frac{\bar{x_{1}} - \bar{x_{2}}}{s_{2} \sqrt{1 + \frac{1}{n_{2}}}}
\end{equation}

where $\bar{x_{1}}$ is the mean within the aperture containing the 
feature, $\bar{x_{2}}$ is the mean of the remaining 6 resolution
elements, $s_{2}$ is the standard deviation of the remaining 6
resolution elements, and $n_{2}$ is the number of resolution elements
not containing the feature.  We find S/N$\sim$6 for the detected
feature, thus this detection is unlikely to be a false positive -- a
result which is further bolstered by the independent detection of the
same feature by Reggiani et al. 2014.

Initially assuming purely photospheric emission, we estimate mass using
Monte Carlo methods to account for the range
of possible ages and uncertainties in photometry for this system.  An
ensemble of 10$^{6}$  possible ages are drawn from a uniform distribution running from 
3-12 Myr.  An ensemble of 10$^{6}$ absolute magnitudes are
simulated, assuming Gaussian errors on photometry and distance (0.3
mag error in $\Delta$mag, 10 pc error in distance).
 We then interpolate with age and single
band absolute magnitude to find the best mass for the companion 
from the DUSTY and COND models of \citet[][]{Bar00} and \citet[][]{Bar03}.    
Mass range histograms are presented in Fig. 2.
Adopting a 3-12 Myr age range at a distance of 145
pc and the DUSTY models, this candidate would correspond to a relatively
high mass 60-80 M$_{Jup}$ brown dwarf.

\subsection{April 2014 Magellan AO Observations}

No counterpart to the L' detection was found in the April 2014
MagAO followup.  Our non-coronagraphic 3.9 $\mu$m followup imaging was
too shallow to retrieve the pointlike feature ($\Delta$mag=5.6 at 0.11'').
However, a 60-80 M$_{Jup}$ object at these ages
should have been quite bright in the near-IR (H band absolute magnitude $<$7) and 
have been easily detected given our H and K$_{s}$ sensitivity.  Indeed, our sensitivity
is sufficient to detect a $\sim$7-15 M$_{Jup}$ object at 
this separation and age range.  Such a high-mass companion should open
up a very wide gap in the disk, inconsistent with the ring at similar
radii observed polarimetrically by Quanz et al. 2013b.
The lack of a near-IR counterpart to the L' detection signifies that we are not observing
the photosphere of a substellar object here but may be a potential disk feature
its extremely red colors indicate that it is extraordinarily unlikely to be a background.

We performed radiative transfer calculations to investigate whether or
not a passively heated disk 'feature' 
can reproduce the observed point-like source.  We used the best fit disk model parameters
from Maaskant et al. 2013 to describe an axisymmetric disk. Then to
simulate a clump we
increased the pressure scale height in the disk. The scale height was increased
as a 2D Gaussian centered on the observed point source with an FWHM of 7\,AU and a peak
perturbation of 30\%. We used 0.1\,$\mu$m size astronomical silicates for the dust opacity
to get an upper limit on the scattered light perturbations. Larger grains would scatter
preferentially forward, perpendicular to the line of sight, which would decrease the contrast
between the clump and the central star. Then we measured the contrast
on the calculated L' images.  We find that a ``dust clump'' at the
position of our candidate would have a contrast of $\sim$17 mag (1$\times$10$^{-7}$)
between the star and the clump, much fainter than the detected candidate.

This result is not surprising, given the distance of the pointlike feature
from the star.
Assuming we observe only thermal dust emission from an irradiated
clump, if the clump peaks at L', it would have T$\sim$850 K.  Thus,
adopting this as the maximum temperature of a potential dust clump, 
an irradiated clump with this temperature would be at a distance of the order of
0.6 AU (considering temperature and radius of the primary from Quanz
et al. 2013b), inconsistent with the $>$10 AU separation of this detection.   
Tentatively, we consider this candidate to be a disk feature
heated by an as yet unknown mechanism.



We also marginally detected a candidate companion in H
and Ks with a separation of $\sim$0.18'' and a PA of
$\sim$33$^{\circ}$, but at low significance (S/N=3-5, 
using the same procedure as for the pointlike
L' feature).   These data were taken at two different
nods, with very different looking PSFs in the two nods,  
limiting the precision of our photometry to $\pm$0.5 mag.    
Thus, while this point source was detected in two
different bands on 3 nights (with S/N$\sim$5 in the first 
dataset, but only S/N$\sim$3 in the additional datasets), 
it requires confirmation at an additional
epoch, with a more stable PSF, to verify common proper motion and rule out imaging and
reduction artifacts.  If real and purely photospheric, this candidate 
would be a 8-15 M$_{Jup}$ planet / 
substellar object (DUSTY and COND models).  No corresponding L'
counterpart was found in the July 2013 L' vortex
dataset.  If purely photospheric, the L' counterpart to this object is expected to have an
absolute magnitude $>$8.5 mag (DUSTY or COND models), corresponding to an 
apparent magnitude of $>$14.3 mag.  This is considerably 
below our achieved contrast for this dataset at this separation 
(limiting magnitude of $\sim$12.7, from insertion and retrieval of PSF
images into the raw data).  We also do not see a counterpart to this
point source in our deep MagAO/VisAO z' dataset,
down to a contrast of 10$^{-4}$.  If real, the very red colors of this
object are inconsistent with background stars with spectral types 
earlier than mid to late M.
 
\section{Discussion}

The known geometry of this disk limits the potential masses of
companions associated with imaged asymmetries.
Quanz et al. 2013b image an inner dust ring peaking at 25 AU and 
an annular gap from 40-70 AU.  
The L' detection lies within the 25 AU dust ring and may be 
connected to ongoing companion formation in the disk.
If associated with a massive object, it may have had some role 
in clearing out the region inside this ring. 
Our tentative H/K detection does not lie within the 40-70 AU gap but 
is coincident with the 25 AU dust ring.  This suggests that any
substellar/planetary counterpart to this tentative detection
cannot be very massive, as a massive object ($\gtrsim$0.55 M$_{Jup}$ 
should completely clear away dust from its immediate environment
(Kley $\&$ Nelson, 2012).  If massive objects are
associated with these detections, they would be some 
of the closest companions imaged to date.  Among confirmed directly imaged
exoplanets, only $\beta$ Pic b, HR 8799d, and HR 8799e have comparably
small projected separations (Lagrange et al. 2010, Marois et al. 2008,
2010).  


Some disk asymmetries detected 
via Sparse Aperture Masking (SAM) have been 
described due to forward scattering off an inclined disk
(e.g. Cieza et al. 2013, Olofsson et al. 2013).   However, HD 169142 is
nearly face on (Quanz et al. 2013b), so forward scattering cannot have 
produced any observable structure in this case.

Our L' detection is also not well-described by dust features heated only by
the central star; it is simply too far from the star to be heated enough
to produce the observed emission.  
However, bright disk-related
features have been observed in other transition disks.  For instance,
Kraus et al. 2013 detect bright asymmetries via sparse aperture
masking in H, K', and L' band within the 15-40 AU 
inner gap in the pre-transitional disk V1247
Orionis.  These asymmetries are not well-fit by a companion model; 
these authors attribute these features to spiral features or
accretion streams due to the interaction of the dust disk with the
substellar bodies inside the gap responsible for its clearing.
Kraus \& Ireland 2012 detect a protoplanet candidate (also via 
sparse aperture masking) around LkCa 15.  While their
detection is well fit by a single point source at K, the L detection
appears extended and not entirely coincident with the K detection. 
They interpret the K detection as the protoplanet itself and the L
detection as circumplanetary material potentially shocked and 
heated by jets from the protoplanet.


In the case of HD 169142, the possible energy source for the observed 
L' disk feature remains unknown.  As previously noted, 
the lack of a H/K counterpart rules out photospheric emission and the
L' detection is too strong to be described by passive heating from the
star.  If due to jets or accretion onto a forming planetary companion, we
would expect a H$\alpha$ counterpart to the L' detection (Close et
al. 2014).  However, Follette et al. in prep have observed HD 169142 with MagAO in H$\alpha$
and do not detect an H$\alpha$ counterpart to the L' disk feature.

The L' detection may be related to emission line regions inside the HD
169142 disk.  Habart et al. 2006 found strong 3.3 $\mu$m PAH emission
(from a C-H stretching feature of neutral PAHs)
in the HD 169142 disk out to 0.3'' (using high-resolution AO-supported
NACO long-slit spectroscopy and along a slit oriented north to south,
i.e. intersecting our detection).  Maaskant et al. 2014 also find
strong PAH emission (however, primarily in ionized features) in the 
HD 169142 disk,  likely stemming from the inner disk gap.
A weak blue leak (on order 1$\%$)
in the NACO L filter might allow us sensitivity to PAH features in the
disk.  If our L' feature is indeed due to PAH emission, future 3.3 $\mu$m 
follow-up observations should easily retrieve this feature.

\section{Conclusions}

We report the detection of a faint pointlike structure in the HD
169142 transition disk, also independently detected by Reggiani 
et al. 2014.  This structure has $\Delta$mag(L)$\sim$6.4, at a separation of
  $\sim$0.11$\arcsec$ and PA of $\sim$0$^{\circ}$.
Given its lack of an H or K$_{S}$ counterpart despite its relative
brightness,
this object cannot be due to the photosphere of a
substellar or planetary mass companion and must instead be a disk feature. 
However, the observed L' detection is too strong to be described by passive heating from the
star and ongoing accretion is ruled out by the MagAO H$\alpha$
non-detection of Follette et al. in prep.  While the location of the
L' pointlike feature right within the disk gap
strongly suggests it may be connected to ongoing planet-formation
(i.e. whatever process cleared out the gap), the energy source fueling this
feature without producing a corresponding near-IR counterpart 
still remains mysterious.  Hopefully, future multi-wavelength followup 
observations will elucidate the source of this feature -- and perhaps
in the process refine our understanding of ongoing planet-formation in this disk.

\acknowledgements

MagAO was constructed with NSF MRI, TSIP, and ATI awards.

\clearpage

\begin{deluxetable}{lcc}
\tabletypesize{\footnotesize}
\tablecaption{Properties of the HD 169142 System\label{tab:properties}}
\tablewidth{0pt}
\tablehead{
\colhead{} & \colhead{Primary} & \colhead{Disk Feature}}

\startdata

Distance      & \multicolumn{2}{c}{145$\pm$10\tablenotemark{a}} \\
Age           & \multicolumn{2}{c}{3-12 Myr\tablenotemark{b}} \\
Proper Motion ($\mu_{\alpha}$, $\mu_{\delta}$) & \multicolumn{2}{c}{(-2.1$\pm$1.5, $-$40.2$\pm$1.5) mas/yr\tablenotemark{c}}  \\

Separation: 14 July 2013 UT      & \nodata  &  0.11$\pm$0.03$\arcsec$ ($\sim$16~AU) \\ 
Position Angle: 14 July 2013 UT   &  \nodata & 0$\pm$14$^{\circ}$ \\


$\Delta L'$ & \nodata & 6.4$\pm$0.3 \\
$H$ (mag)                        & 6.91\tablenotemark{d} & \nodata \\
$Ks$ (mag)                       & 6.41\tablenotemark{d} & \nodata \\
$L'$ (mag)                       &  5.64 \tablenotemark{e} &  12.0$\pm$0.3 \\
$M_{L'}$ (mag)                   & -0.2 & 6.2$\pm$0.3 \\ 
Spectral type                    &  A7V$\tablenotemark{f}$   &\nodata \\
Estimated Mass   & 1.65 M$_{\odot}$\tablenotemark{f} & \nodata \\
\enddata
\tablenotetext{a}{\citet[][]{Syl96}, but errors derived here.}
\tablenotetext{b}{\citet[][]{Gra07}}
\tablenotetext{c}{\citet[][]{Hog00}}
\tablenotetext{d}{From 2MASS}
\tablenotetext{e}{\citet[][]{Mal98}}
\tablenotetext{f}{\citet[][]{Blo06}}

\end{deluxetable}

\begin{deluxetable}{lcccc}
\tabletypesize{\footnotesize}
\tablecaption{Observation log \label{tab:observations}}
\tablewidth{0pt}
\tablehead{
\colhead{Telescope/Instrument} & \colhead{UT Date} & \colhead{Band}
&\colhead{Exposure Time} & \colhead{Field Rotation}}
\startdata
VLT NACO Vortex & 14 July 2013 & L' & 75 min & 135.2 \\
\hline 
MagAO CLIO-2 Narrow & 9 April 2014 & H & 46 min & 178.5 \\
MagAO CLIO-2 Narrow & 10 April 2014 & H & 30.4 min & 172.9 \\
MagAO CLIO-2 Narrow & 12 April 2014 & K$_{S}$ & 70 min & 180.1 \\
MagAO CLIO-2 Narrow & 15 April 2014 & 3.9$\mu$m & 60 min & 164.1 \\
MagAO VisAO Narrow & 15 April 2014 & z$_{p}$ & 82.4 & 176.8 \\
\enddata 
\end{deluxetable}

\begin{figure}
\includegraphics[width=4in]{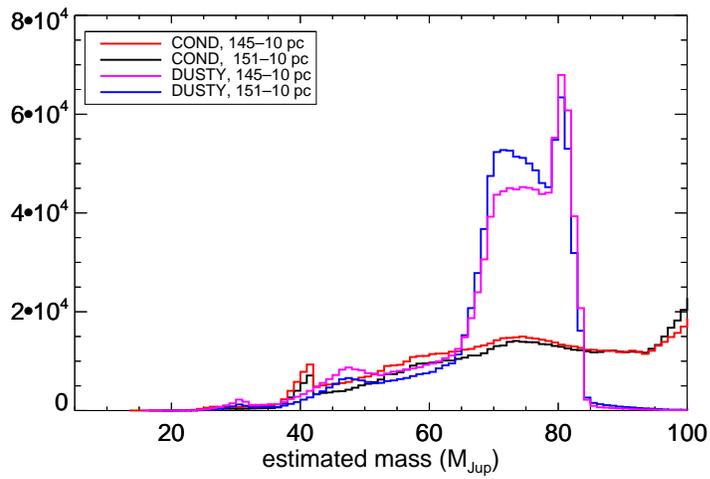} 
\caption{Mass estimate histogram for the pointlike feature observed
  in L'. If due to photospheric emission from a 
companion, this object would have mass $>$40 M$_{Jup}$ and 
an easily detected H and Ks counterpart.  
\label{fig:masshist}
}
\end{figure}




\clearpage

\end{document}